\documentclass[conference]{IEEEtran}
\IEEEoverridecommandlockouts

\usepackage{tikz}
\usepackage{float}
\usepackage{forest}
\usetikzlibrary{arrows.meta, shapes.geometric, positioning, fit, backgrounds}
\usepackage{booktabs}
\usepackage{array}
\usepackage{natbib}
\usepackage{amsmath,amssymb,amsfonts}
\usepackage{algorithmic}
\usepackage{graphicx}
\usepackage{textcomp}
\usepackage{xcolor}
\usepackage{makecell}

\usepackage{hyperref}
\hypersetup{
    colorlinks=true,
    linkcolor=darkblue,
    citecolor=darkblue,
    urlcolor=darkblue
}
\definecolor{darkblue}{rgb}{0.0, 0.0, 0.55}

\def\BibTeX{{\rm B\kern-.05em{\sc i\kern-.025em b}\kern-.08em T\kern-.1667em\lower.7ex\hbox{E}\kern-.125emX}}
\begin{document}

\title{
Passive Acoustic Monitoring of Underwater Well Leakages with Machine Learning: A Review
%
\thanks{\hspace*{-\parindent}\rule{3.8cm}{0.4pt} \\ 
$\ast$: Equal contribution. \\
$\dagger$: Corresponding author.}
}
\author{\IEEEauthorblockN{Guanlin Zhu$^{\ast}$}
\IEEEauthorblockA{\textit{School of Physics \& Astronomy} \\
\textit{University of Edinburgh}\\
Edinburgh, United Kingdom\\
G.Zhu-10\\
@sms.ed.ac.uk}

\and

\IEEEauthorblockN{Zechun Deng$^{\ast}$}
\IEEEauthorblockA{\textit{School of Physics \& Astronomy} \\
\textit{University of Edinburgh}\\
Edinburgh, United Kingdom\\
Z.Deng-16\\
@sms.ed.ac.uk}

\and

\IEEEauthorblockN{Jiaxin Shen}
\IEEEauthorblockA{\textit{Wolfson School}\\
\textit{Loughborough University}\\
London, United Kingdom\\
J.Shen5-19\\
@student.lboro.ac.uk}

\and

\IEEEauthorblockN{Junchi Yang}
\IEEEauthorblockA{\textit{School of Physics \& Astronomy} \\
\textit{University of Glasgow}\\
Glasgow, United Kingdom\\
3075465Y\\
@student.gla.ac.uk}
}



\maketitle

\begin{abstract}
Abandoned oil and gas wells pose significant environmental risks due to the potential leakage of hydrocarbons, brine and chemical pollutants. Detecting such leaks remains extremely challenging due to the weak acoustic emission and high ambient noise in the deep sea. This paper reviews the application of passive sonar systems combined with artificial intelligence (AI) in underwater oil and gas leak detection. The advantages and limitations of traditional monitoring methods, including fibre optic, capacitive and pH sensors, are compared with those of passive sonar systems. Advanced AI methods that enhance signal discrimination, noise suppression and data interpretation capabilities are explored for leak detection. Emerging solutions such as embedded AI analogue-to-digital converters (ADCs), deep learning-based denoising networks and semantically aware underwater optical communication (UOC) frameworks are also discussed to overcome issues such as low signal-to-noise ratio (SNR) and transmission instability. Furthermore, a hybrid approach combining non-negative matrix factorisation (NMF), convolutional neural networks (CNN) and temporal models (GRU, TCN) is proposed to improve the classification and quantification accuracy of leak events. Despite challenges such as data scarcity and environmental change, AI-assisted passive sonar has shown great potential in real-time, energy-efficient and non-invasive underwater monitoring, contributing to sustainable environmental protection and maritime safety management.
\end{abstract}

\begin{IEEEkeywords}
Passive sonar, subsea leak detection, artificial intelligence (AI), machine learning, underwater optical communication (UOC), deep learning, non-negative matrix factorisation (NMF), digital signal processing,  environmental monitoring.
\end{IEEEkeywords}

\section{Introduction and Motivation}
\begin{figure*}
    \centering
    \includegraphics[width=0.96\linewidth]{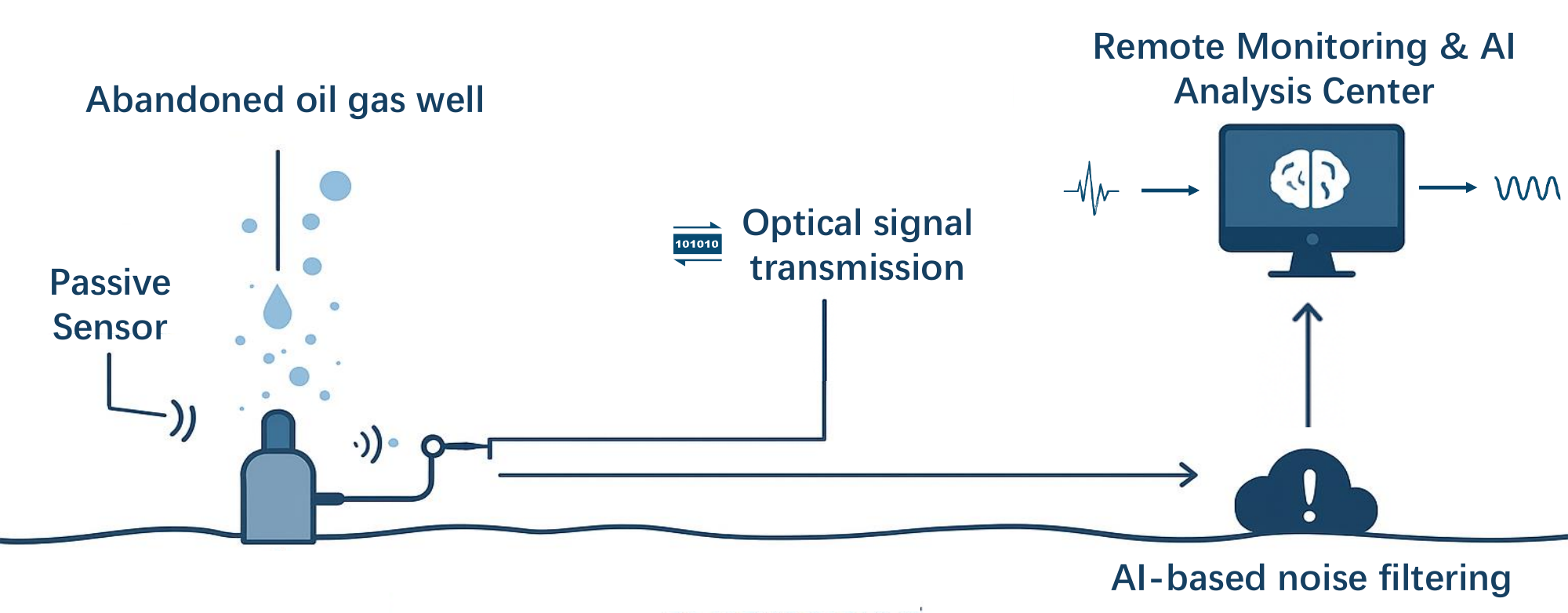}
    \caption{\textbf{AI-enhanced sensor system for subsea oil and gas leakage detection.}}
    \label{fig:intro}
\end{figure*}
After cessation of oil and gas production, wells are typically abandoned in accordance with regulatory and environmental requirements. In the United States, there are more than two million abandoned oil and gas wells \cite{intro1}. Due to environmental effects and creature activities, abandoned wells tend to leak gas, oil and other chemicals, which can seriously contaminate the environment and threaten the health of humans and animals \cite{intro2}. However, current underwater monitoring technologies often struggle to detect oil and gas leaks accurately and promptly due to ambient noise and harsh working conditions. Therefore, it is critical to find a reliable method to diagnose the leakage for timely maintenance and pollution abatement accurately.

Given the complicated surroundings and vulnerable pipelines, non-invasive detecting technologies, such as acoustic monitoring, fibre optics, chemical sensors, etc., are much more suitable for monitoring the status of abandoned oil and gas wells. Compared to other technologies, passive sonar does not require deploying on the surface of the infrastructure or introducing an additional signal source to capture oil and gas leakage. This technology allows for the monitoring of non-destructive and environmentally friendly leaks of underwater wells \cite{intro3}. Nevertheless, the signal-to-noise ratio (SNR) of passive technology tends to be low due to the complicated ambient noise and the weak signal of oil and gas leakage. That means it is difficult for the monitoring system to diagnose the slight leakage early.

At the same time, because the seabed is far away from the sea surface, the data transmission process will also be affected by noise in the ocean, such as the noise of marine life \cite{intro4}. This requires digital signal processing at the sensor and during data transmission to filter and reduce noise as much as possible. To prevent signal dissipation during transmission in the ocean, Underwater Optical Communication (UOC) would be a better choice than other transmissions, such as underwater electric communication or underwater acoustic communication (UAC) \cite{intro5}. After the terminal collects the required data, the leaked oil and gas data will be classified by algorithms. For the collected data sets, the images can be pre-processed, such as image enhancement, noise reduction and filtering \cite{intro6}. After that, the collected data can be classified through cluster analysis and other methods.

AI techniques can be deployed at every stage in the sensor system. Adding AI can help passive sonar better identify small leaks. Furthermore, AI can be used during data transmission to reduce interference from other ocean noise. AI technology can also be applied to filtering, image processing and signal enhancement \cite{intro7}. In short, the usage of AI technology can significantly improve the accuracy of subsea oil and gas leakage detection.

\section{Current Methods to Detect Oil and Gas Leakage from Abandoned Wellhead} 

The oil and gas sector constitutes a critical component of the global economy. An estimated 10 million barrels of crude oil are extracted daily, and more than 2 million oil and gas wells are abandoned in the USA alone \cite{intro1, intro6}. Hence, oil and gas well leakage is one of the most significant problems to be solved due to its serious environmental impact and dramatic economic loss. Various advanced technologies have been developed for monitoring leakages, including approaches based on acoustics, optics, imaging, chemistry, biology and geophysics \cite{method_intro}. This section provides a brief analysis of the advantages of the passive acoustic approach over alternative monitoring technologies, including optical fibre sensing, capacitive sensing and pH sensing techniques.

According to the principle of refraction, when the incident angle of the light reaches and exceeds a critical value, the transmitted ray ceases to propagate into the lower-refractive-index medium. Instead, it travels along the boundary between the two media. That means the wave is completely reflected within the medium with a higher refractive index. Based on the characteristics of light, such as phase, intensity, and wavelength. The optical fibre sensor can be deployed to detect leaks from underwater oil and gas wells. \cite{opticfibre1} found that several problems exist with the present optic fibres, such as short sensing and frequency range, high ambient interference and high cost, among others.

A capacitor consists of two electrodes separated by an insulating material, known as a dielectric. When voltage is applied, charge accumulates on the electrodes until the electric field reaches equilibrium. The capacitance depends on the dielectric properties, plate separation and electrode area \cite{capa2}. In seabed oil and gas leakage detection, capacitive sensors can be placed near wellheads or structures, where hydrocarbons entering seawater or sediment alter the local dielectric constant and thus the measured capacitance. Because hydrocarbons have a much lower dielectric constant than seawater, even small leaks can be detected in real time \cite{capa1}. Capacitive sensors are passive, compact and fast-responding. Still, their deep-water use faces challenges: maintaining electrode geometry under pressure, avoiding parasitic capacitance in conductive seawater and preventing biofouling that reduces sensitivity. Their limited detection range and the small dielectric contrast between hydrocarbons and saltwater also reduce the SNR \cite{capa3}. Thus, effective deep-sea applications demand careful sensor design with protective electrodes, shielding, calibration, and anti-fouling coatings.

\begin{table*}[ht]
\centering
\renewcommand\cellalign{tl}
\renewcommand\cellgape{\Gape[2.5pt]}
\caption{\textbf{Comparison of Subsea Oil and Gas Leak Detection Techniques.}}
\begin{tabular}{|>{\raggedright\arraybackslash}p{4.5cm} 
                |>{\raggedright\arraybackslash}p{5.7cm} 
                |>{\raggedright\arraybackslash}p{5.7cm}|}
\hline
\multicolumn{1}{|c|}{\textbf{Technology}} &
\multicolumn{1}{c|}{\textbf{Advantages}} &
\multicolumn{1}{c|}{\textbf{Limitations}} \\
\hline
\raisebox{-0.5em} {Optical Fibre Sensing} & \raisebox{-0.5em} {High sensitivity} & \makecell [l] {Limited sensing range, High cost, \\Environmentally sensitive}\\
\hline
\raisebox{-0.5em} {Capacitive Sensing} & \makecell [l] {No moving parts, Compact structure, \\Fast response} & \makecell [l] {Biofouling issues, Limited sensing range, \\Parasitic capacitance interference}\\
\hline
\raisebox{-0.5em} {pH Sensing} & \makecell [l] {High sensitivity, Fast response, \\Continuous monitoring, No moving parts} & \makecell [l] {Signal interference, Long-term signal drift, \\Environmentally sensitive} \\
\hline
\raisebox{-0.5em} {Passive Acoustic Monitoring System} & \makecell [l] {Non-invasive, Low energy consumption, \\Continuous real-time monitoring} & \makecell [l] {Noise interference, Localization difficulty, \\Environmental dependence}\\
\hline
\end{tabular}
\end{table*}

Fluorescence-based pH sensors offer significant advantages for monitoring oil and gas leaks from abandoned subsea wellheads, particularly due to their high sensitivity, rapid response, lack of moving parts and continuous, in situ, real-time measurement \cite{pH1}. Furthermore, pH sensors exhibit minimal drift over short periods of time (e.g., for continuous fluorescence pH monitoring in seawater), facilitating deployment in harsh deep-sea environments \cite{pH2}. In addition to detecting pH perturbations caused by oil/gas leaks in the seabed, they face several challenges. Interference from naturally occurring fluorescent substances in seawater (e.g. chlorophyll and humic compounds) can corrupt the pH signal. Signal drift and photobleaching over long deployments require frequent calibration or drift correction. Biofouling on the sensor surface can reduce sensitivity. Additionally, extreme pressure, temperature gradients, salinity, and heterogeneous multiphase fluids near leaking wells can complicate optical access and stability of the fluorescent environment \cite{pH3}. In this case, fluorescence-based pH sensing could be powerful for high-resolution detection of acidified patches surrounding leakage. However, careful design (e.g. antifouling coatings, in situ calibration, and optical shielding) is required to overcome the limitations of multiphase leakage environments in the deep ocean.

The passive sonar system integrates hydrophones as sensing devices to receive and analyse the acoustic waves emitted from oil and gas leak sources. The passive sonar system can identify leakage events by analysing the frequency spectrum of acoustic signals produced by the escape of fluids and gases from the pipeline \cite{method_passivesonar1}. In comparison with alternative monitoring methods, passive sonar demonstrates distinct advantages such as its non-invasive nature, low energy consumption, stealthy detection capability and suitability for continuous real-time monitoring of underwater leakage events. Although passive sonar is an effective tool for leakage monitoring, its performance can be hindered by several factors. These include interference from ambient noise, difficulties in accurately localising and quantifying leakage sources, inherent sensitivity limitations and a strong dependence on environmental parameters such as temperature, salinity and background flow \cite{method_passivesonar2}. Moreover, reliable detection requires sophisticated signal processing and regular system calibration.

Underwater oil and gas leak detection employs multiple sensor technologies, each with distinct principles and limitations. Optical fibre sensors detect leaks via light refraction and phase or intensity changes, but they face short sensing ranges, high interference, and cost issues \cite{opticfibre1}. Capacitive sensors monitor local dielectric changes caused by hydrocarbons, offering real-time detection. However, deep-sea pressure, bio-fouling and limited range pose challenges \cite{capa2}. Fluorescence-based pH sensors offer sensitive, continuous monitoring of acidification in the vicinity of leaks. Yet interference, photobleaching and harsh multiphase conditions require careful calibration \cite{pH1}. Passive sonar systems analyse sounds generated by fluid and gas. Passive acoustic monitoring (PAM) offers a promising non-invasive approach for detecting underwater oil and gas leaks due to its adaptability and real-time capabilities. However, the ambient noise, localisation difficulties and environmental dependence affect the accuracy \cite{method_passivesonar2}.
\section{Advanced AI Methodologies for Passive Sonar Applications}
Artificial intelligence (AI) technologies have become essential tools for enhancing the performance of passive sonar systems used in detecting subsea oil and gas leaks. As mentioned, passive sonar systems face significant challenges, including difficulties in localising and quantifying leakage sources, sensitivity limitations and strong ambient interferences. AI can be effectively applied to address these issues. This section analyses over thirty related studies and explores three key domains of AI application: methods for differentiating sensor signals, techniques for noise reduction, disturbance mitigation and approaches for processing and interpreting collected data. By integrating AI into passive sonar systems across these areas, the detection accuracy, robustness, and efficiency of subsea monitoring can be substantially improved.

\subsection{Methods for Differentiating Sensor Signals}
\begin{figure*}
    \centering
    \includegraphics[width=0.88\linewidth]{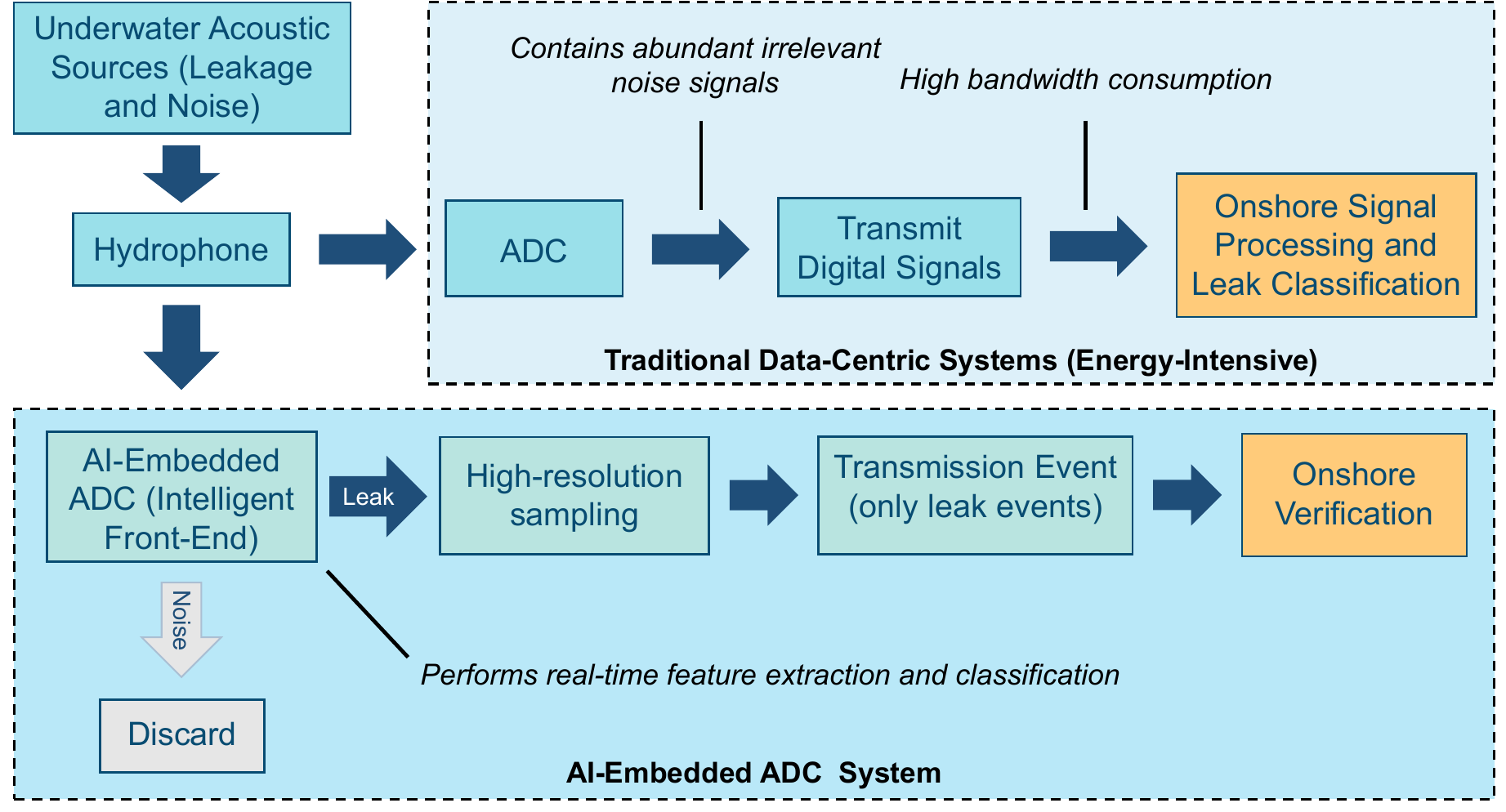}
    \caption{\textbf{Paradigm shift: from data-centred to semantic-centred.}}
    \label{fig:signal}
\end{figure*}

Subsea leaks primarily radiate sound through the formation and collapse of bubbles. This sound wave spectrum is typically broadband, with peaks related to bubble size and flow conditions (similar to Minnaert behaviour). Gas plumes tend to concentrate their energy in the tens of kHz and above; passive sonar typically targets leak detection in the 30-150 kHz range \cite{signal1}. Furthermore, sea conditions, ship and bioacoustics are major confounding factors during detection, while these signals have varying time and frequency signatures \cite{signal2}. The application of AI can immediately distinguish interfering signals from detected leak signals that need to be transmitted.

Traditional sonar and sensor systems rely on digitising all acoustic data at the analogue-to-digital converter ($\Delta$ - $\Sigma$ ADC) and performing classification downstream \cite{signal3}. While this approach maintains signal fidelity, it results in high energy consumption, bandwidth usage and overload with irrelevant background noise (e.g., acoustic noise from shipping, biological sources, and weather \cite{signal4}). Literature on underwater signal classification demonstrates the effectiveness of feature-based machine learning and deep neural networks for post-processing \cite{signal5}. Embedding AI directly at the ADC level to classify and transmit detected signals can improve ADC efficiency and energy consumption \cite{signal6}. Current research trends include spectral and statistical feature extraction, anomaly detection models \cite{signal7} and self-supervised learning methods for addressing limited labels, all of which can enhance leak detection performance when applied offline or higher up the signal chain \cite{signal8}. However, processing methods must transmit or store raw data before AI-based classification presents significant limitations. Power-constrained subsea platforms cannot sustain high-rate transmission, most digitised samples have little semantic value for leak detection \cite{signal9}.

\cite{signal10} therefore, propose a novel perspective: the ADC can be redefined as an intelligent front-end rather than a passive digitiser. By embedding a lightweight AI model within or near the converter, the system can distinguish oil and gas leakage signals from irrelevant noise in real time and selectively transmit only meaningful events. This aligns with the goal-oriented semantic communication framework, which aims to minimise transmitted information while preserving task-relevant inference accuracy. Formalised as $\min\limits_{g(\cdot)}\space H(\widetilde{X})$, subject to $\widetilde{X}=g(X)$ and $I(\widetilde{X};Y)\geq I(X;Y)$ \cite{signal14}. Where $g(\cdot)$ represents the semantic feature extraction performed at the converter, $H(\widetilde{X})$ represents the minimum required transmission rate, $\widetilde{X}$ is the compressed task-relevant representation, and $Y$ is the decision variable.

This approach combines the extraction of mixed signal characteristics (such as envelope detectors, frequency band energy ratios and modulation signatures) with a compact neural classifier \cite{signal11} to perform semantic filtering before the data leave the sensor. This allows the system to distinguish the oil and gas leak signal from irrelevant background noise. When a potential leak is detected with high confidence, the ADC triggers high-resolution sampling and transmits the event. Otherwise, it discards or compresses irrelevant data, significantly reducing bandwidth and energy consumption \cite{signal12}. Reliability is ensured through fallback modes, audit sampling and self-calibration, while interpretability is maintained by logging features and decisions.

By embedding AI in the ADC pipeline, it optimises data capture for leak detection tasks, not just signal fidelity \cite{signal13}. This represents a paradigm shift from data-centric sensing ("collect all data and analyse it later") to semantic sensing ("collect only the important data and analyse it early") \cite{signal5}. This paradigm reduces bandwidth and energy requirements while improving detection reliability.
\subsection{Methods for Noise Reduction and Disturbance Mitigation}
\begin{figure*}
    \centering
    \includegraphics[width=0.88\linewidth]{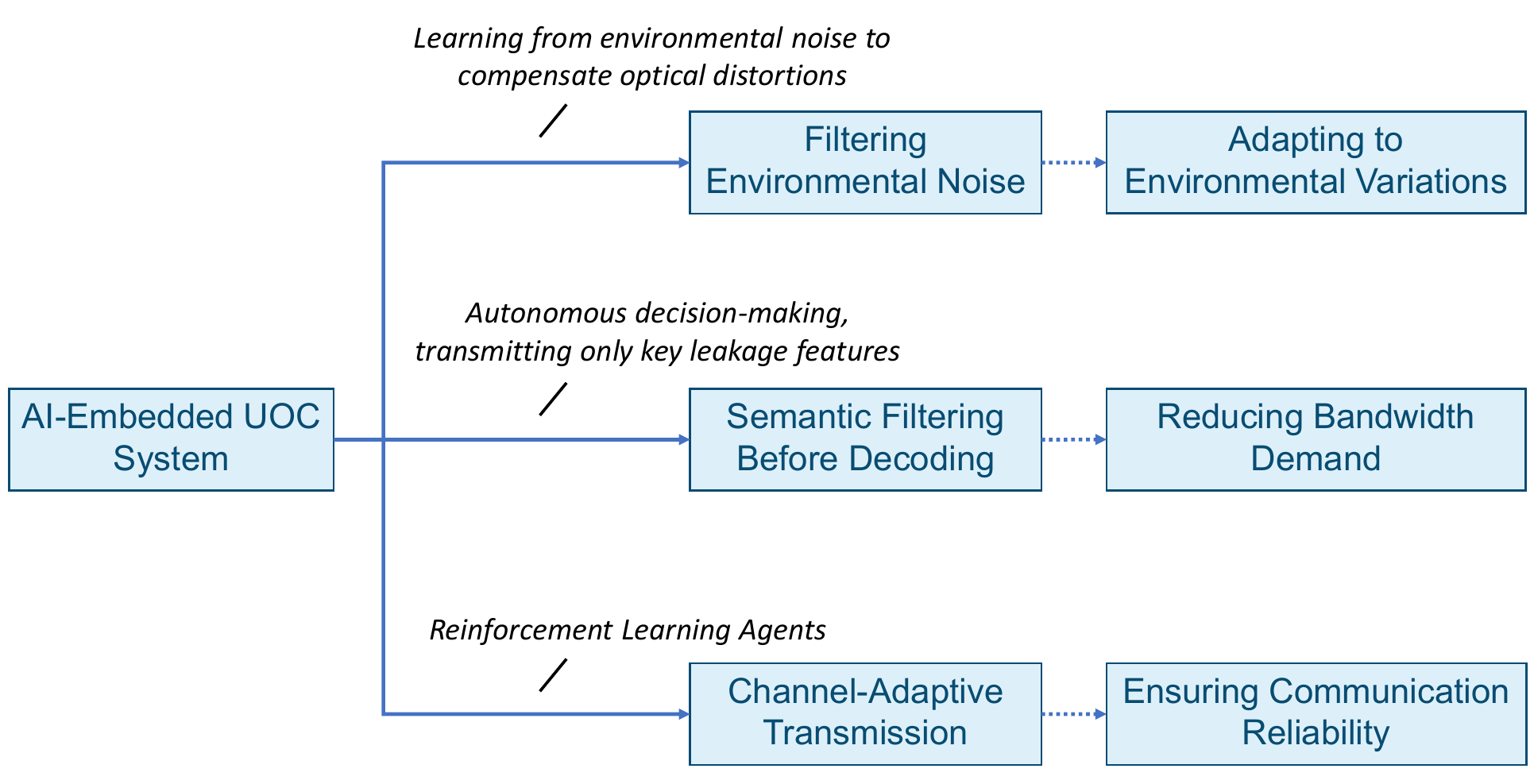}
    \caption{\textbf{AI-embedded Underwater Optical Communication (UOC) system and its functional pathways.}}
    \label{fig:DSP}
\end{figure*}

The use of passive sonar to detect fluid or gas leaks in underwater pipelines and infrastructure is increasingly challenged by environmental noise, human interference and propagation distortions, as pointed out by \cite{dsp8} (e.g., wind, rain, marine life, ship noise). Artificial intelligence techniques, particularly machine learning (ML) and deep learning (DL), have been proven by \cite{dsp9} to provide a foundation for improving the signal-to-noise ratio (SNR), suppressing interference and enhancing the robustness and operational readiness of passive systems \cite{dsp10}. \cite{dsp11} indicates the growing application of AI methods in underwater acoustic signal enhancement, denoising and target recognition. Although much research focuses on target recognition, like ships or marine life \cite{dsp12}. The underlying technology is highly relevant to leak detection systems, as the "target" of a leak detection system is the weak acoustic emission generated by the leak.

\cite{dsp7} explicitly addresses the acoustic denoising problem in underwater remote sensing: a "dual-stream deep learning-based acoustic denoising" model is proposed to suppress environmental noise in passive sonar applications, thereby making low-frequency target signals more prominent. This can be directly applied to leak monitoring, as weak low-frequency radiation can be masked by environmental noise. \cite{dsp13} found one branch of the network that estimates a spectral mask $M(f)$ while the other models temporal correlations of noise. The two outputs are fused to minimise the reconstruction loss between the clean target $s(t)$ and the enhanced signals $s^t$, as shown in the formula \ref{eq1}.
\begin{equation}
    \mathcal{L} = \frac{1}{T} \sum_{t=1}^{T} \left( s(t) - \hat{s}(t) \right)^{2}.
    \label{eq1}
\end{equation}
To complement the passive sonar system, underwater optical communication (UOC) can serve as a high-capacity backhaul link to transmit leakage detection data from the seabed to surface terminals \cite{intro5}. However, UOC links are notoriously sensitive to scattering, absorption, turbulence and biofouling, which degrade signal quality and increase error rates \cite{dsp1}. Embedding AI directly in the UOC transceiver offers a pathway to mitigate these impairments at the physical layer \cite{dsp2}. Lightweight neural equalisers, deployed in the transmitter and receiver pipeline, can be optimised with channel coding to learn the non-linear distortions induced by seawater turbulence and particulate scattering \cite{dsp3}. Edge-deployed models can run in near real-time, either on embedded processors within the optical modem or at the photodetector interface, to perform semantic filtering before high-level decoding \cite{dsp2}. Unlike conventional DSP-based equalisers, AI-enabled modules can dynamically adapt to local water conditions and compensate optical distortions, thus stabilising the UOC link even in challenging environments \cite{dsp4}.

Deploying AI in UOC can earn a good performance in denoising and semantic error correction, rather than treating all distortions as equivalent. First, as \cite{dsp3} said, deep generative models can reconstruct clean signal waveforms by learning the statistical structure of modulated optical signals in the presence of noise, effectively serving as a “neural optical filter.” Second, semantic-aware joint source channel coding can be implemented, where an auto-encoder learns to transmit only features essential for leak detection (e.g., event confidence, spectral fingerprints), reducing bandwidth needs while maintaining detection reliability \cite{dsp5}. Third, reinforcement learning agents can be employed for adaptive beam steering and power allocation, adjusting transmission parameters in response to channel fluctuations. Together, these AI techniques redefine UOC as an intelligent communication system \cite{dsp6}. Not only transporting data but also enhancing, filtering and prioritising information before it leaves the seabed node. This integration of AI with UOC thus represents a paradigm shift from raw data relay to intelligent, resilient and task-optimised subsea communication.
\subsection{Methods for Processing and Interpreting Data}
\begin{figure*}
    \centering
    \includegraphics[width=0.88\linewidth]{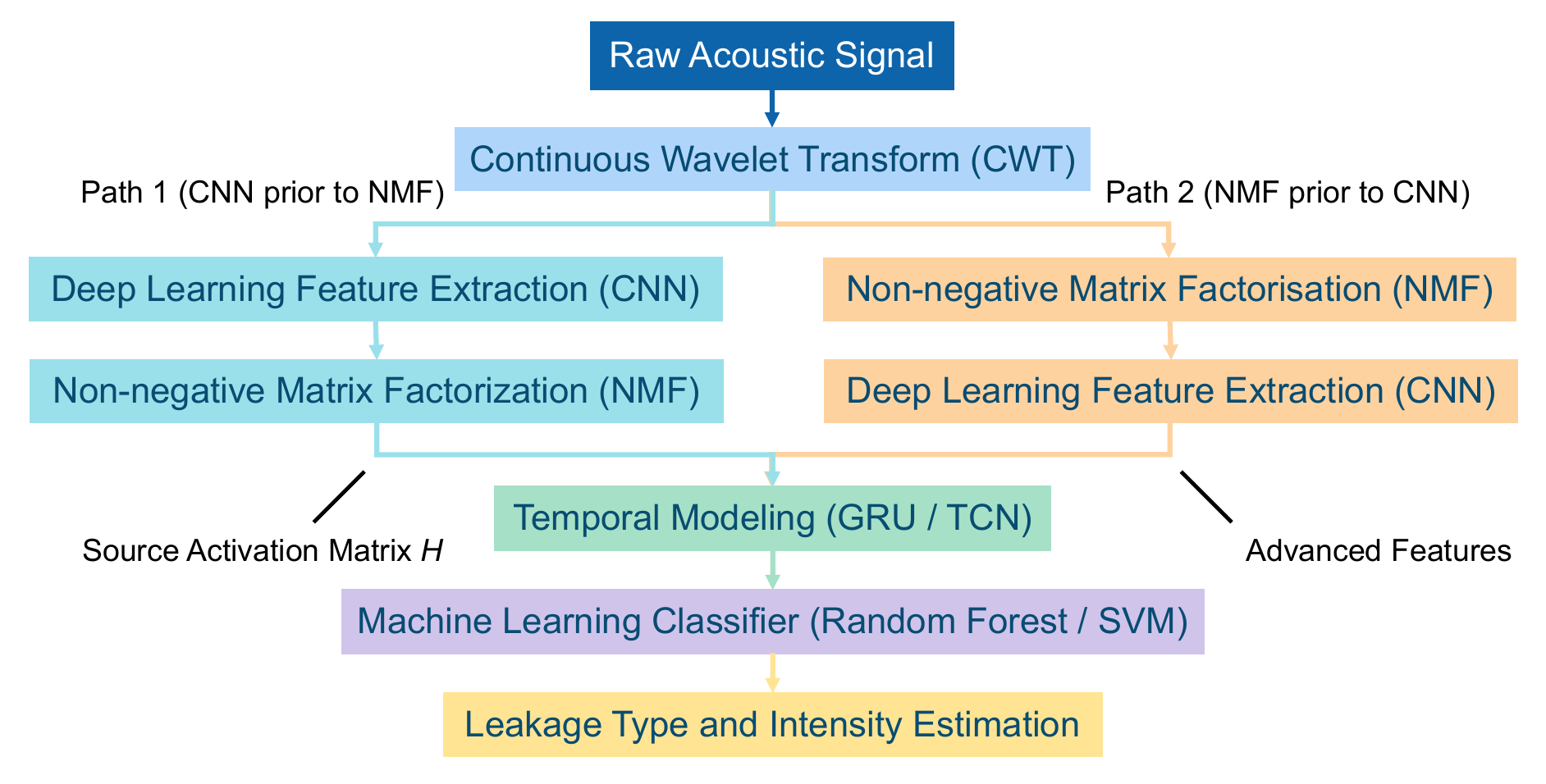}
    \caption{\textbf{Conceptual framework of using AI techniques to analyse the collected underwater acoustic data.}}
    \label{fig:analysis}
\end{figure*}

The underwater leakage signals are easily disturbed by strong noise from marine life and water waves \cite{intro4}, resulting in a low signal-to-noise ratio (SNR). The Fourier transform is an effective way to analyse the signal \cite{ana1}, which converts the time-domain signal $x(t)$ to the frequency-domain:
\begin{equation}
    X(f) = \int^{\infty}_{-\infty} x(t) e^{-j2\pi ft}dt.
\end{equation}
This transformation enables analysis of the frequency content of the signal as pointed out by \cite{ana2}, allowing engineers to locate spectral peaks that correspond to a leakage event. However, there are some limitations, as the Fourier transform assumes signal stationarity, which is often impossible in real-world scenarios \cite{ana3}. Based on the research of \cite{ana4} to overcome this problem, the Continuous Wavelet Transform (CWT) can help capture time and frequency information simultaneously:
\begin{equation}
    W(a,b) = \frac{1}{\sqrt{|a|}}\int^{\infty}_{-\infty}x(t)\psi^*(\frac{t-b}{a})dt,
\end{equation}
Where $x(t)$ is the recorded acoustic signal, $\psi(t)$ is the mother wavelet, $a$ is the scale parameter, and $b$ is the time shift parameter. This architecture led to better results in an unstable underwater environment \cite{ana5}, thereby improving the robustness and interpretability of leak signal analysis. In this stage, signal classification and leakage intensity quantification have been covered, which approximate the flow rate \cite{ana6}. The Non-negative Matrix Factorisation (NMF) technique satisfies both source separation and quantification by decomposing the observed spectrogram $V$ into basis patterns $W$ and corresponding temporal activations $H$.
\begin{equation}
V \approx W * H,
\end{equation}
Where $V \in \mathbb{R}^{F\times T} _{+} $ is the signal spectrogram, $ W \in \mathbb{R}^{F \times K}_{+} $ for oil and gas leak patterns contains spectral bases and $H \in \mathbb{R}^{K\times T}_{+}$ gives the activation varying over time of each component. Training $W_{oil}$ and $W_{gas}$ in controlled recordings allows the system to estimate $H$ for unseen mixed data \cite{ana7}. In this way, the relative contributions of oil and gas leakage will be quantified and this makes NMF both interpretable and physically meaningful for subsea monitoring applications \cite{ana8}.

To enhance the robustness and accuracy in a noisy subsea environment, a supervised Machine Learning method will be adopted \cite{cha2}, which combines NMF outputs ($H$) with classifiers or regressors, such as Random Forest and Support Vector Machines (SVMs), for improved discrimination. Found by \cite{ana9}, convolutional Neural Networks (CNNs) can be applied before or after NMF to automatically extract discriminative spectral-spatial features from the decomposed spectrograms, thereby comprehending the non-linear relationships. Gated Recurrent Units (GRU) and Temporal Convolutional Networks (TCN) can provide an efficient solution for temporal modelling \cite{ana10}. To help track how the oil and gas leakage intensities change over time. Where GRUs can capture long-term dependencies in the activation matrix $H$ with fewer parameters. TCNs can be used to model time correlations in parallel in causal convolutions \cite{ana11}. This improved the effectiveness and stability of the training. These two models help identify the different leakage behaviours under varying subsea conditions.

After combining NMF and supervised ML classifiers, the robustness under noise will be improved. Deep learning, such as CNNs and RNNs, can also act as feature extractors before or after NMF decomposition \cite{ana5}, which shows features from spectrograms. Time-series models, e.g. GRU, TCN, can further capture dynamic changes in $H$. As mentioned in \cite{ana10}, allowing the system to track how leakage evolves. By adopting multi-task architectures, the probability of oil and gas, as well as the estimated intensity, can be output from $H$ or the regression head \cite{ana12}. By applying NMF and machine learning algorithms, leakage detection can become more accurate in identifying a specific number of oil and gas leaks.


\section{Potential Challenges and Strategies}
\begin{figure*}
    \centering
    \includegraphics[width=0.88\linewidth]{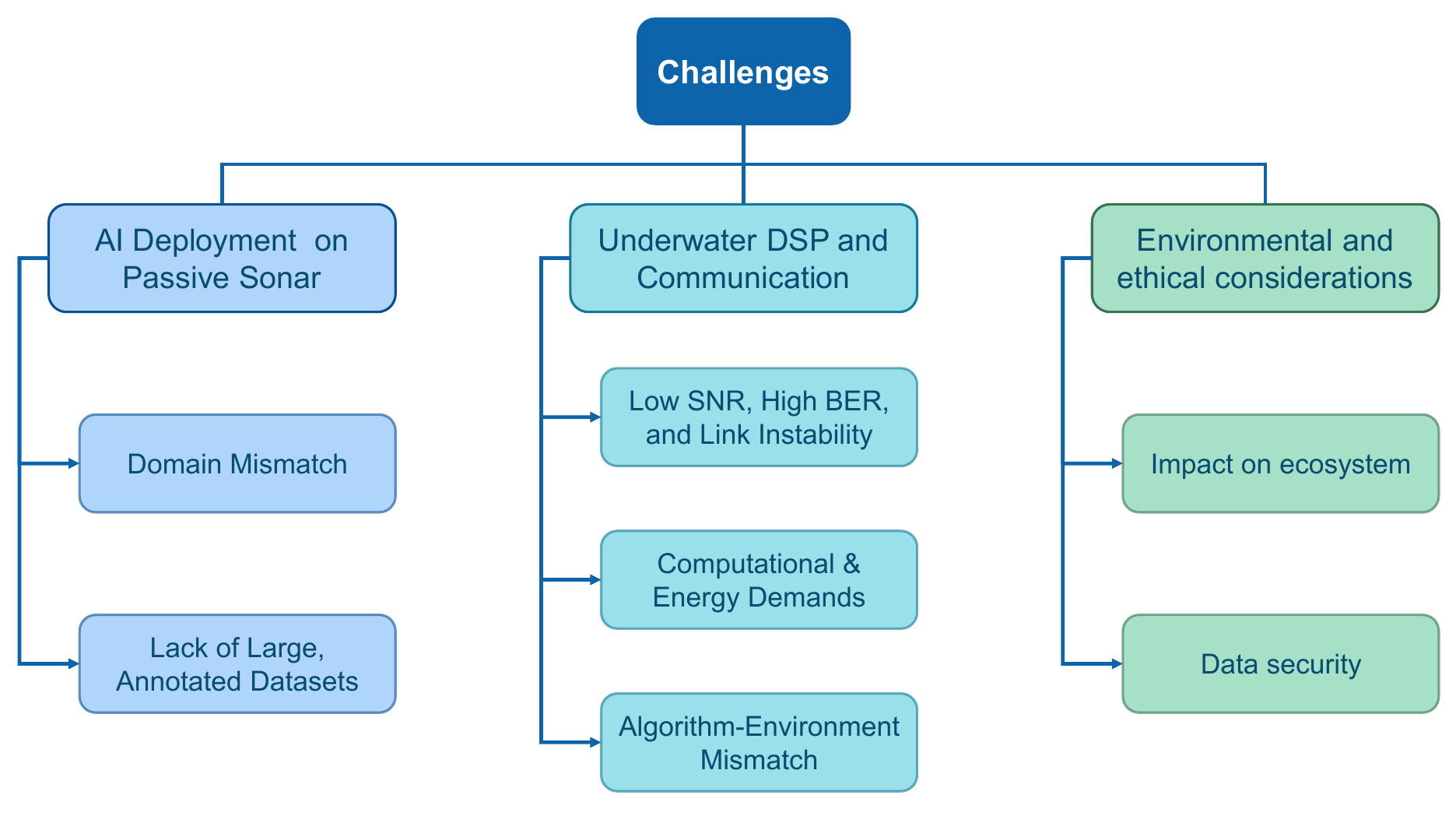}
    \caption{\textbf{Overview of technical and environmental challenges in AI-based passive sonar.}}
    \label{fig:challenge}
\end{figure*}

A key barrier to the deployment of AI in passive sonar leakage detection is the scarcity of large, annotated underwater acoustic datasets, especially for leak signals. \cite{cha2} emphasises that within sonar data classification, there is a severe lack of public datasets, which hinders reproducibility and generalisation. In addition, the problem of domain mismatch is acute. As \cite{cha1} pointed out, sensors, geometry, bathymetry, sea state, shipping traffic and noise conditions vary widely from one deployment to another. Moreover, AI models may be too close to training conditions and perform poorly in new ones \cite{cha3}. To overcome current shortages, the strategies proposed in the article, based on \cite{cha4}, include the addition of informed physical data, the simulation of real transfer learning, self-supervised pretraining and continuous adaptive learning.

In acoustic channels, long propagation delays and Doppler shifts distort time coherence, leading to inter-symbol interference (ISI) and performance degradation of phase-sensitive modulation schemes \cite{cha5}. In optical communication, \cite{cha6} noticed that the absorption of water molecules and the scattering of suspended particles severely limit the transmission distance and stability of the link, while water turbulence causes random intensity fluctuations, resulting in signal fading \cite{cha7}. These challenges collectively contribute to low signal-to-noise ratio (SNR), high bit error rate (BER) and link instability in underwater environments. Furthermore, in the research of \cite{cha8}, high computational complexity and energy consumption make many traditional DSP techniques unsuitable for resource-constrained underwater sensor nodes. The mismatch between algorithm adaptability and environmental changes remains one of the key obstacles to achieving reliable underwater communication.

A potential research direction by \cite{cha9} is to jointly optimise digital signal processing (DSP) and communication protocols using reinforcement learning (RL) and semantic communication frameworks. RL-based controllers can dynamically allocate transmit power, modulation schemes and beam directions based on real-time feedback from channel states \cite{cha10}, thereby optimising link reliability under constantly changing ocean conditions. Meanwhile, \cite{cha11} found that semantically aware source channel coding enables the system to prioritise the transmission of task-related information (such as leak event confidence or frequency patterns) rather than uniformly transmitting all data. This task-oriented transmission approach reduces bandwidth and energy consumption, aligning with the concept of "intelligent edge awareness." 

Although significant progress has been made by applying an AI algorithm to subsea leak detection, there are still several challenges to be addressed. The low SNR characteristic of underwater environments is one of the problems. Marine animals, vessel traffic, and water often produce noise that masks the signal collection \cite{intro4}. Based on \cite{cha13}, future studies can focus on developing methods and noise-efficient architectures to achieve stable performance across different environments. Environmental and ethical considerations are also important and should be addressed in the future. The placement of sensors and detectors underwater can affect the ecosystem \cite{cha14}. Eco-friendly materials and responsible data management should be considered to help align technology with the environment.

Finally, combine with the findings of \cite{cha12}, a hybrid solution with traditional digital signal processing (DSP) for deterministic processing with artificial intelligence (AI) models for adaptive inference ensures reliability, interpretability and computational efficiency in practical underwater deployments.

\section{Conclusion}
This review explores the potential of AI-enhanced passive acoustic monitoring (PAM) technology for detecting oil and gas leaks at abandoned subsea wellheads. Compared to fibre optic, capacitive and pH sensing technologies, passive sonar offers a non-invasive, energy-efficient and real-time solution, particularly suitable for harsh and complex underwater environments \cite{method_passivesonar2}. However, its practical application remains limited by environmental noise interference, location uncertainty and low signal-to-noise ratio (SNR).

The integration of AI technologies offers some insights into overcoming these limitations. AI-embedded analogue-to-digital converters (ADCs) enable real-time signal classification and event-triggered data transmission, shifting the sensing paradigm from data-centric to semantic-centric. Similarly, deep learning-based denoising models \cite{dsp9} and AI-enhanced underwater optical communication (UOC) frameworks \cite{con2} improve the reliability of data acquisition and transmission under turbulent ocean conditions. Regarding data interpretation, combining nonnegative matrix factorisation (NMF) with machine learning and temporal neural networks (e.g., GRU, TCN) enables accurate quantification of leak type and intensity under diverse seabed conditions \cite{signal5}. Despite these advances, several challenges remain, particularly the scarcity of labelled underwater acoustic datasets, domain mismatch and computational limitations of deep-sea sensor nodes \cite{cha2}. Future research should prioritise RL, self-supervised training and reinforcement learning-based optimisation to enhance system adaptability and robustness \cite{con1}.

In summary, AI-assisted passive sonar systems represent a promising direction for sustainable underwater monitoring, capable of improving the accuracy, efficiency and autonomy of detecting leaks at abandoned wellheads. Collaboration among acoustics, signal processing and machine learning is crucial for translating these advancements into reliable, field-deployed systems. These systems could support environmental protection and energy security management in the global maritime industry in the future.

\bibliographystyle{apalike}
\bibliography{ref}

\end{document}